\documentclass{PoS}

\title{The spectrum of radial, orbital and gluonic excitations of charmonium}
\ShortTitle{The spectrum of radial, orbital and gluonic excitations of 
charmonium}
\author{K.J.~Juge\\ 
  Department of Physics Carnegie Mellon University, Pittsburgh, USA.}
\author{A.~\'O Cais, \speaker{M.B. Oktay}. M.J.~Peardon, 
        S.M.~Ryan, J-I.~Skullerud\\
        School of Mathematics, Trinity College, Dublin, Ireland\\
        Email:\email{oktay@maths.tcd.ie}
       }

\rightline{\sffamily TRINLAT-06/10}

\abstract{
We present results for the charmonium spectrum from $N_f=2$ dynamical QCD 
simulations on $12^3\times 80$ anisotropic lattices. Using all-to-all 
propagators we determine the ground and excited states of S, P and D waves 
and hybrids. We also evaluate the disconnected (OZI suppressed) contribution 
to the $\eta_c$ and $J/\Psi$.
}

\FullConference{XXIV International Symposium on Lattice Field Theory\\
		 July 23-28, 2006\\
		 Tucson, Arizona, USA}

\begin{document}

\section{Introduction}

In recent years there has been a resurgence of interest in charmonium 
physics, both theoretically and experimentally. 
The charmonium states below threshold are considered important topics for
study by lattice QCD. They provide us with a testing ground for lattice 
heavy  quark methods 
which are also used to determine elements of the CKM matrix involving $b$ 
quarks. 
New experiments at CLEO-c aim to confront lattice QCD calculations with 
extremely precise 
experimental data from the charm sector. In addition, the discovery of
 numerous new, and unexplained, 
bound states such as the $X(3871)$ and the presence of theoretically allowed 
exotic states has renewed efforts in charm physics. 

Lattice QCD can in principle answer many of the outstanding questions but to 
do so requires 
high precision numerical simulations. 

In these proceedings we present preliminary results from a relativistic 
simulation of the charmonium spectrum, using all-to-all propagators on a 
dynamical anisotropic lattice. An earlier study of the spectrum on a smaller 
volume was presented in 
Ref.~\cite{Juge:2005nr}. The charmonium spectrum at finite temperature 
was also determined in this way and is presented in 
Refs~\cite{Allton:2006pos,Aarts:2006nn,Morrin:2005zq}. 

\section{The dynamical anisotropic action}

The gauge action is a two-plaquette Symanzik-improved action. It is written 
\begin{equation}
S_{G} = \frac{\beta}{\xi^0_g} \left\{ \frac{5(1+\omega)}
         {3u^4_s}\Omega_s - \frac{5\omega}{3u^8_s}\Omega^{(2t)}_s - 
\frac{1}{ 12u^6_s} \Omega^{(R)}_s \right\} 
         +\beta\xi^0_g\left\{ \frac{4}{3u^2_s u^2_t} \Omega_{t} - 
	 \frac{1}{12u^4_s u^2_t} \Omega^{(R)}_t \right \}
\end{equation}
where $\Omega_{s,t}^{R,2t}$ refer to simple and rectangular plaquettes in
the spatial and temporal directions and $\xi_g^0$ is the bare anisotropy. 
The full details are described in Ref~\cite{Morningstar:1999dh}. 

The anisotropic fermion action~\cite{Foley:2004jf} is written
\begin{equation}
S_q = \bar{\Psi}\left[ \gamma_0\nabla_0+\sum_i\mu_r\gamma_i\nabla_i\left(
         1-\frac{1}{\xi_q^0a_s^2}\Delta_i\right) -\frac{ra_t}{2}\Delta_{i0} 
    + \frac{sa_s^3}\sum_i\Delta_i^2 +m_0\right]\Psi .
\end{equation}
where $a_s$ and $a_t$ ($a_s\gg a_t$) are the spatial and temporal lattice 
spacings respectively and $\xi_q^0$ is the bare anisotropy.  

In an anisotropic simulation the bare anisotropies, which are inputs in the 
gauge and fermion 
actions must be tuned so that the measured value of the anisotropy in a 
simulation takes its ``target'' value. In the quenched theory the gauge and 
fermion  anisotropies can be separately tuned: however in the dynamical 
theory these must be simultaneously tuned. The details of this tuning for 
quenched and dynamical QCD are described in 
Refs.~\cite{Foley:2004jf,Morrin:2005tc,Morrin:2006tf}. 

\section{Simulation details}
We performed simulations at $N_f=2$ on $N_s\times N_t = 12^3\times 80$ 
lattices with 250 configurations at a sea quark mass close to the strange 
quark mass. The simulation details are summarized in Table~\ref{tab:details}.
\begin{table}[h]
  \begin{center}
    \begin{tabular}{ll}
      \hline
      Configurations & 250 ($a_tm_c=0.117, a_tm_{\rm sea}
                             =a_tm_{\rm light}=-0.057$) \\
      Dilution       & time and space even/odd     \\
      Physics        & S,P,D waves, S wave radial excitations,
      and hybrids        \\
      Volume         & $12^3\times 80$    \\
      $N_f$          & 2  \\
      $a_s$          & $\sim$ 0.17 fm \\
      $a_t^{-1}$     & $\sim$ 7GeV \\
      $m_\pi/m_\rho$ & $\sim 0.55$ \\
      \hline
    \end{tabular}
    \caption{Simulation details for this study of charmonium. }
    \label{tab:details}
  \end{center}
\end{table}
The target (renormalised) anisotropy in this study is 6. We found that, 
in contrast to the quenched case~\cite{Foley:2004jf} the anisotropy in the 
charm sector had 
to be tuned separately to that in the light quark sector to obtain the same 
renormalised 
anisotropy. This mass-dependence of $\xi$ in the dynamical theory may be a 
discretisation effect and is under investigation. The charm quark mass used 
in the simulation, 
$a_tm_c=0.117$, was tuned to obtain the correct $J/\Psi$ mass. We 
use all-to-all propagators with dilution, no eigenvectors and two noise
vectors, as described in Ref.~\cite{Foley:2005ac}, 
for better signal to noise ratios. An additional advantage of all-to-all 
propagators is that  constructing an extended basis of operators for better 
overlap with the states of interest is easier than with point propagators. 
The list of operators used in this study is given in 
Table~\ref{tab:operators}.
\begin{table}[h]
  \begin{center}
    \begin{tabular}{lc|lc}
      \hline
      $0^{-+}$ & $\gamma_5$,$\gamma_5(s_1+s_2+s_3)$ & $2^{++}$ 
      & $\gamma_k p_i+\gamma_i p_k$  \\
      $1^{--}$ & $\vec{\gamma}$,$\gamma_j(s_1+s_2+s_3)$ & $2^{-+}$ 
      & $\gamma_5(2s_3-s_1-s_3)$ \\
      $1^{+-}$ & $\gamma_5\vec{p}$ & $2^{--}$ & $\gamma_j(s_i-s_k)$ \\
      $0^{++}$ & $\vec{\gamma}\cdot\vec{p}$ & $3^{--}$ 
      & $\vec{\gamma}\cdot\vec{t}$\\
      $1^{++}$ & $\vec{\gamma}\times\vec{p}$ & $1^{-+}$ 
      & $\vec{\gamma}\times\vec{u}$ \\
      \hline
    \end{tabular}
    \caption{The basis of operators. The notation used for the gluonic 
      paths, denoted with $s_i, i=1,2,3$, is defined in 
      Ref.~\cite{Lacock:1996vy}.}
    \label{tab:operators}
  \end{center}
\end{table}
The spin-averaged (1P-1S) splitting is used to set the lattice spacing, 
yielding $a_t=0.028$fm and $a_s=0.17$fm. 

\begin{figure}[h]
  \begin{center}
    \includegraphics*[scale=0.4]{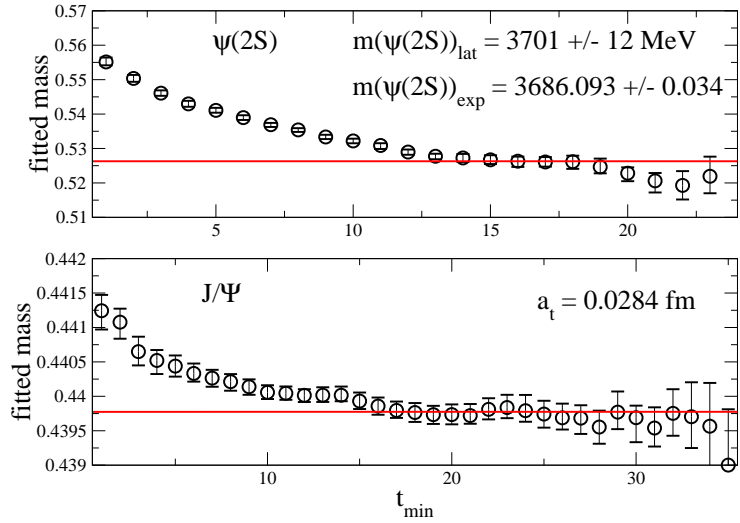}
  \end{center}
  \caption{Sliding window plots for the ground and first excited 
    states of the $J/\psi$ at zero momentum. The solid lines show the final
 best-fitted masses chosen.}
    \label{fig:slide}
\end{figure}

Since time dilution introduces a random noise source at each timeslice,
effective masses fluctuate more from timeslice to timeslice than is usually
seen with point propagators making it difficult to identify a good plateau 
region. However, these fluctuations do not affect exponential fits and can 
be reduced further by increasing the dilution level. 
A more accurate reflection of the quality of the data and stability of the 
fits can be seen from a sliding window plot. For a fixed value of 
$t_{\rm max}$ the fitting window, $(t_{\rm min}, t_{\rm max})$ is varied by 
changing $t_{\rm min}$. The sliding window plots show the fitted masses at
each interval as a function of $t_{\rm min}$. Sliding window plots for 1S 
and 2S states of $J/\psi$ are shown in Figure~\ref{fig:slide}. 
The plots show that the 1S fitted mass is stable as $t_{\rm min}$ is varied 
over twenty timeslices and the 2S mass is stable over seven values of 
$t_{\rm min}$. For all the fitted masses shown here the 
quality of the fit is determined and the stable regions also have good 
$\chi^2/N_{\rm d.f.}$. The other charmonium states studied also show 
extremely stable fitted masses. Based on this analysis the hyperfine 
splitting is approximately 30 MeV in contrast to the experimental value 
of 117MeV. The reason for this underestimate is not known yet and will be 
investigated in future work. In particular, the lattice spacing in this 
study is quite coarse and the action does not include a $\Sigma\cdot B$ 
term. In addition it is known from previous work that the hyperfine 
splitting increases with decreasing lattice spacing and for lighter sea 
quark masses in dynamical simulations. These improvements will be included 
in future simulations. 

\begin{figure}[h]
  \begin{center}
    \includegraphics*[scale=0.4]{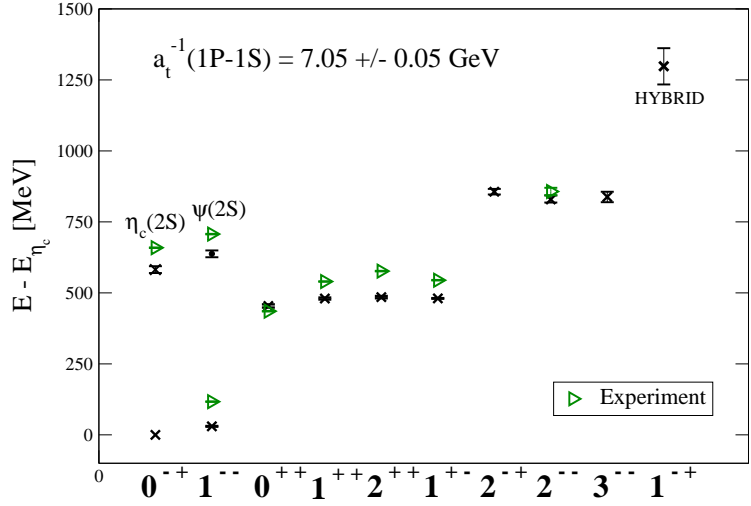}
  \end{center}
  \caption{The $c\bar{c}$ spectrum which is normalised to the $\eta_c$ mass.
    The scale has been set from the $(1P-1\bar{S})$ splitting. Experimental 
values, where known, are denoted 
with right-facing triangles. }
    \label{fig:spectrum}
\end{figure}

To determine the radially excited states we use a variational basis of 
operators included extended operators. Our results for the ground state 
mass using this approach agrees within two percent with the same state 
determined from exponential fits. Two smearings and two different operators 
were used to increase the variational basis when determining the $\eta_c$ 
and $J/\Psi$. The first excited state of the 
$\eta_c$ and $J/\Psi$ are $m_{\eta_c(2S)}=3645\pm12$ MeV and 
$m_{\Psi (2S)}=3701\pm 12$ MeV respectively. This 
analysis is being extended to the P and D waves. Our preliminary charmonium 
spectrum, including the P-waves, D-waves and the $1^{-+}$ hybrid is shown 
in Figure~\ref{fig:spectrum}.

\section{Disconnected diagrams}
\begin{figure}[h]
  \begin{center}
    \includegraphics*[scale=0.4]{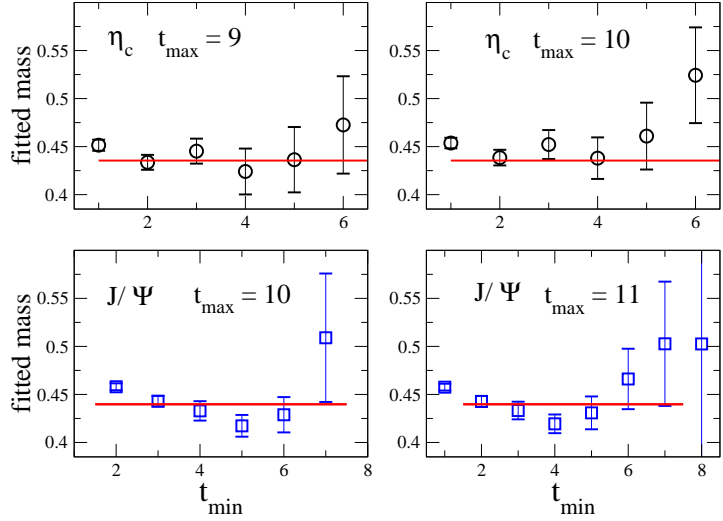}
    \caption{Sliding window plots showing the fitted mass in lattice units 
      for the full correlators, 
      $C_{full}=C_{conn}-C_{dis}$ for the $\eta_c$ and $J/\psi$. 
      The plots show the effect of 
      changing $t_{\rm max}$ in the fits. The poor quality of the signal 
      means $t_{\rm max}$ has been 
      moved in from the centre of the lattice approximately timeslice ten. 
      The straight line represents the ground state mass obtained from 
      the connected parts of the correlators.}
    \label{fig:ozi}
  \end{center}
\end{figure}

In most lattice calculations of the charmonium spectrum and the hyperfine
splitting in particular, the effects of disconnected diagrams in the two 
point correlator are ignored. It has been estimated that these 
OZI-suppressed diagrams may in fact contribute $\sim$ 20 MeV to the 
hyperfine splitting~\cite{McNeile:2004wu,deForcrand:2004ia}, 
bringing traditionally low lattice determinations into better agreement with 
experiment. However, the inclusion of such diagrams is not easy, requiring 
all-to-all propagators which in the past have resulted in noisy data 
from which it is difficult to extract a convincing signal. It is hoped that 
the all-to-all propagator algorithm used here may solve this problem using
the appropriate dilution to reduce noise. In this first attempt the diagrams 
were directly calculated using the same dilution already described 
(time and space). We found that the disconnected diagrams are still noisy 
and in fits to the full correlator, given by the difference of the 
connected and disconnected correlators the fit window is reduced to 
ten data points. Sliding window plots for the full correlators, 
$C_{\rm full}(t) = C_{\rm connected}(t)-C_{\rm disconnected}(t))$ 
are shown in figure \ref{fig:ozi}. 
The plots show no discernible difference after including the disconnected 
contributions, however we are currently investigating other dilution 
schemes to optimise the signal. 

\section{Conclusions}
We have presented our preliminary results from a simulation of the charmonium
spectrum on $N_f=2$ dynamical $12^3\times80$ anisotropic lattices. 
All-to-all propagators and a variational basis of operators have been
used the determine the spectrum of S, P and D waves, their radial excitations
as well as the $1^{-+}$ hybrid with good statistical precision. 
In addition, we have investigated the effects of the disconnected diagrams
on the hyperfine splitting. We find a negligible difference. However, 
these preliminary results need to be investigated further. In particular 
we are planning to explore dilution strategies and study the volume 
dependence of the higher excited states. 

\section*{Acknowledgements}
This work was supported by the IITAC project, funded by 
the Irish Higher Education Authority under PRTLI 
cycle 3 of the National Development Plan and funded by SFI grant 
04/BRG/P0275 and IRCSET grant SC/03/393Y.

\bibliography{oktaymb_lat06.bib}

\end{document}